\documentclass[aps,prl,showpacs,twocolumn,groupedaddress]{revtex4}  

\usepackage{graphicx}  
\usepackage{dcolumn}   
\usepackage{bm}        
\usepackage{amssymb}   
\usepackage{d0_style}

\def\lsim{\mathrel{\rlap{\lower4pt\hbox{\hskip1pt$\sim$}}\raise1pt\hbox{$<$}}}
\def\gsim{\mathrel{\rlap{\lower4pt\hbox{\hskip1pt$\sim$}}\raise1pt\hbox{$>$}}}

\begin{document}

\hspace{5.1in} \mbox{FERMILAB-PUB-09-046-E}

\title{Measurement of the {\boldmath$Z\gamma \to \nu\bar\nu\gamma$} cross section and limits on anomalous {\boldmath$ZZ\gamma$} 
and {\boldmath$Z\gamma\gamma$} couplings in {\boldmath$p\bar{p}$} collisions at {\boldmath$\sqrt{s}$}~=~1.96~TeV}

%
\author{V.M.~Abazov$^{36}$}
\author{B.~Abbott$^{75}$}
\author{M.~Abolins$^{65}$}
\author{B.S.~Acharya$^{29}$}
\author{M.~Adams$^{51}$}
\author{T.~Adams$^{49}$}
\author{E.~Aguilo$^{6}$}
\author{M.~Ahsan$^{59}$}
\author{G.D.~Alexeev$^{36}$}
\author{G.~Alkhazov$^{40}$}
% ...

\affiliation{\vspace{0.1 in}(The D\O\ Collaboration)\vspace{0.1 in}}
% ...
\affiliation{$^{6}$University of Alberta, Edmonton, Alberta, Canada,
                Simon Fraser University, Burnaby, British Columbia, Canada,
                York University, Toronto, Ontario, Canada, and
                McGill University, Montreal, Quebec, Canada}
% ...
\affiliation{$^{29}$Tata Institute of Fundamental Research, Mumbai, India}
% ...
\affiliation{$^{36}$Joint Institute for Nuclear Research, Dubna, Russia}
% ...
\affiliation{$^{40}$Petersburg Nuclear Physics Institute,
                St. Petersburg, Russia}
% ...
\affiliation{$^{49}$Florida State University, Tallahassee, Florida 32306, USA}
% ...
\affiliation{$^{51}$University of Illinois at Chicago,
                Chicago, Illinois 60607, USA}
% ...
\affiliation{$^{59}$Kansas State University, Manhattan, Kansas 66506, USA}
% ...
\affiliation{$^{65}$Michigan State University,
                East Lansing, Michigan 48824, USA}
% ...
\affiliation{$^{75}$University of Oklahoma, Norman, Oklahoma 73019, USA}
% ...

\begin{abstract}
We present the first observation of the $Z\gamma \to \nu\bar\nu\gamma$ process at the Tevatron 
at 5.1 standard deviations significance, based on 3.6~fb$^{-1}$ of integrated luminosity collected with the D0~detector 
at the Fermilab Tevatron $p\bar{p}$ Collider at $\sqrt{s}$ = 1.96 TeV. The measured $Z\gamma$ cross section 
multiplied by the branching fraction of $Z\to\nu\bar\nu$ is $32 \pm 9 {\rm (stat.+syst.)} \pm 2 {\rm (lumi.)~fb}$ 
for the photon $E_T > 90$~GeV. It is in agreement with the standard model prediction of $39\pm4$~fb. 
We set the most restrictive limits on anomalous trilinear $Z\gamma\gamma$ and $ZZ\gamma$ gauge boson couplings at a 
hadron collider to date, with three constraints being the world's strongest.
\end{abstract}

\maketitle

The standard model (SM) of electroweak interactions is described by the
non-Abelian gauge group $SU(2)\times U(1)$. The symmetry transformations 
of the group allow interactions involving three gauge bosons ($\gamma$, $W$, and $Z$) 
through trilinear gauge boson couplings. However, 
the SM forbids such vertices for the photon and the $Z$ boson 
at the lowest ``tree'' level, $i.e.$, the values of the $Z\gamma\gamma$ and $ZZ\gamma$ 
couplings vanish. The cross section for the SM $Z\gamma$ production is very small. 
However, the presence of finite ({\it{anomalous}}) 
$Z\gamma\gamma$ and $ZZ\gamma$ couplings can enhance the yields, especially at 
higher values of the photon transverse energy ($E_{T}$). 
As we are marginally sensitive to one-loop SM contributions~\cite{Gounaris_2002za,zzg-theory} to 
these vertices, observation of an anomalously high $Z\gamma$ production rate could, therefore, 
indicate the presence of new physics.

To preserve $S$-matrix unitarity, the anomalous couplings must vanish at high 
center-of-mass energies. Hence, the dependence on the 
center-of-mass energy has to be included in the definition of such couplings. 
This can be done by using a set of eight complex parameters 
$h^{V}_{i} (i=1,...,4; V = Z, \gamma)$ of the form $h_{i}^{V} = h_{i0}^{V}/(1+\hat s/\Lambda^{2})^{n}$
~\cite{baur}.
Here, $\hat s$ is the square of the center-of-mass energy in the partonic collision, $\Lambda$ is a scale 
related to the mass of the new physics responsible for anomalous $Z\gamma$ production, 
and $h^{V}_{i0}$ is the low energy approximation of the coupling. Following  
Ref.~\cite{baur}, we will use $n = 3$ for $h^{V}_{1}$ and $h^{V}_{3}$, and $n = 4$ 
for $h^{V}_{2}$ and $h^{V}_{4}$. This choice of $n$ guarantees the preservation of
partial-wave unitarity, and makes the vertex function terms proportional to $h^{V}_{1}$ and $h^{V}_{3}$ 
behave in the same way as terms proportional to $h^{V}_{2}$ and $h^{V}_{4}$ at high energies. 
Couplings $h^{V}_{10}$ and $h^{V}_{20}$ ($h^{V}_{30}$
and $h^{V}_{40}$) are CP-violating (CP-conserving). In this Letter, we set limits on the size of 
the real parts of the anomalous couplings: $Re(h^{V}_{i0})$, which we refer to as ATGC in the 
following. 

In the past, studies of $Z\gamma$ production have been performed by the CDF~\cite{cdf} and 
D0~\cite{d01, zg_p17} collaborations at the Tevatron Collider, as well as at the CERN LEP Collider by the 
L3~\cite{l3}, and OPAL~\cite{opal} collaborations. The most recent combination of LEP results can be found in 
Ref.~\cite{lep}. 

The D0~detector~\cite{run2det} consists of a central-tracking system, liquid-argon/uranium
calorimeters, and a muon system. The tracking system comprises a silicon microstrip tracker 
(SMT) and a central fiber tracker (CFT), both located within a $\approx$~2~T superconducting 
solenoid, and provides tracking and vertexing up to pseudorapidities~\cite{eta} of $|\eta| \approx$~3.0 
and $|\eta|\approx$~2.5, respectively. The central and forward preshower detectors (CPS and FPS) 
are located between the superconducting coil and the calorimeters, and consist of three and four 
layers of scintillator strips, respectively. The liquid-argon/uranium calorimeter is divided into 
a central calorimeter (CC) and two end calorimeters (EC), covering pseudorapidities up to 
$|\eta|\approx$~1.1 and $|\eta|\approx$~4.2, respectively. The calorimeters are segmented 
into an electromagnetic section (EM), comprised of four layers, and a hadronic section, divided longitudinally 
into fine and coarse sections. The calorimeter is followed by the muon system,  
consisting of three layers of tracking detectors and scintillation trigger counters and a 1.8~T iron toroidal 
magnet located between the two innermost layers. The muon system provides coverage to $|\eta|\approx$~2. 
Arrays of plastic scintillators in front of the EC cryostats are used to measure the luminosity.

Data for this analysis were collected with the D0 detector in the period from 2002 to 2008, and 
correspond to an integrated luminosity of 3.6~fb$^{-1}$ after the application of data-quality 
and trigger requirements. Events must satisfy a trigger from a set of high-$E_T$ single EM-cluster 
triggers, which are $(99 \pm 1)\%$ efficient for photons of $E_T >$~90~GeV.

Photons are identified as calorimeter clusters with at least $95\%$ of their energy 
deposited in the EM calorimeter, with transverse and longitudinal distributions consistent
with those of a photon, and spatially isolated in the calorimeter and in the
tracker. A cluster is isolated in the calorimeter if the isolation variable 
${\cal{I}} = [E_{\text{tot}}(0.4)-E_{\text{EM}}(0.2)]/E_{\text{EM}}(0.2)~<~0.07$. 
Here, $E_{\text{tot}}(0.4)$ is the total energy (corrected for the contribution from multiple 
$p\bar p$ interactions) deposited in a calorimeter cone of radius 
${\cal{R}}=\sqrt{(\Delta\eta)^{2}+(\Delta\phi)^{2}} = 0.4$, 
and $E_{\text{EM}}(0.2)$ is the EM energy in a cone of radius ${\cal{R}}=0.2$. The track 
isolation variable, defined as the scalar sum of the transverse momenta of all tracks that 
originate from the interaction vertex in an annulus of $0.05 < {\cal{R}} < 0.4$ around 
the cluster, must be less than 2~GeV.

We obtain the photon sample by selecting events with a single 
photon candidate of $E_{T}~>~90$~GeV and $|\eta| < 1.1$, and require a missing transverse 
energy in the event of $\met >$~70~GeV, which effectively suppresses the multijet background. 
The $\met$ is computed as the negative vector sum of the $E_{T}$ of calorimeter cells and corrected 
for the transverse momentum of reconstructed muons and the energy corrections to reconstructed electrons 
and jets. To minimize large
$\met$ from mismeasurement of jet energy, we reject events with jets with
$E_{T} > 15$~GeV. We also reject events containing reconstructed muons, and events with cosmic-ray muons
identified through a timing of their signal in the muon scintillators. 
Events with additional EM objects with $E_{T} > 15$~GeV are rejected. To suppress $W$ boson decays 
into leptons, events with reconstructed high-$p_{T}$ tracks are removed. To reduce the copious non-collision background 
(events in which muons from the beam halo or cosmic rays undergo bremsstrahlung, and produce energetic photons), 
we use a pointing algorithm~\cite{point}, exploiting the transverse and longitudinal energy distribution in the 
EM calorimeter and CPS. This algorithm is based on estimates of $z$ positions of production vertices ($z_{\text{EM}}$) 
along the beam direction assuming that given EM showers are initiated by photons,  
and utilizes the distance of closest approach (DCA)~\cite{dca} of the direction of the EM shower 
to the $z$ axis. We require $|z_{\text{EM}} -z_{\text{V}}| < 10$~cm, where $z_{\text{EM}}$ is the $z$
position of the interaction vertex predicted by the pointing algorithm and
$z_{\text{V}}$ is the $z$ position of the chosen (often nearest) reconstructed vertex.

Following the procedure described in Ref.~\cite{led}, we estimate the fraction of non-collision and 
$W/Z$ events with misidentified jets backgrounds in the final candidate events by fitting their 
DCA distribution to a linear sum of three DCA templates. These templates are: a template resembling the signal, 
a non-collision template, and a misidentified jets template. Most of the signal photons are concentrated in the 
region with DCA~$<$~4~cm. Therefore, we restrict the analysis to this particular range.

Other backgrounds to the $\gamma + \met$ signal 
arise from electroweak processes such as $W \to e\nu$, where
the electron is misidentified as a photon due to inefficiency of the
tracker or hard bremsstrahlung, and $W\gamma$, where the lepton from
the $W$ boson decay is not reconstructed. 

The $W\to e\nu$ background is estimated using a sample of isolated
electrons. We apply the same kinematic requirements as in the photon
sample, and scale the remaining number of events by the measured rate of
electron-photon misidentification, which is $0.014 \pm 0.001$. 
The $W\gamma$ background is estimated using 
a sample of Monte Carlo (MC) events generated with {\sc pythia}~\cite{pythia}. 
These events are passed through a detector simulation chain based on
the {\sc geant} package~\cite{geant}, and reconstructed using the same software
as used for data. After imposing the same selection requirements as for the
photon sample, scale factors are applied to correct for differences between
simulation and data. The summary of backgrounds is shown in Table~\ref{tab:bkg}.

\begin{table}
\caption{Summary of background estimates, and the number of observed and SM predicted events.\label{tab:bkg}}
\begin{ruledtabular}
\begin{tabular}{lr}
              & Number of events \\ \hline
$W \to e\nu$            & $9.67 \pm 0.30(\rm stat.) \pm 0.48(\rm syst.)$\\
non-collision		& $5.33 \pm 0.39(\rm stat.) \pm 1.91(\rm syst.)$\\
$W/Z$ + jet             & $1.37 \pm 0.26(\rm stat.) \pm 0.91(\rm syst.)$\\
$W\gamma$		& $0.90 \pm 0.07(\rm stat.) \pm 0.12(\rm syst.)$\\ \hline
Total background	& $17.3 \pm 0.6(\rm stat.) \pm 2.3(\rm syst.)$\\
$N_{\nu\bar\nu\gamma}^{\rm SM}$ & $33.7 \pm 3.4$\\\hline
$N_{\rm obs}$		& 51 \\
\end{tabular}
\end{ruledtabular}
\end{table}

After applying all selection criteria, we observe 51 candidate events 
with a predicted background of $17.3~\pm~0.6{\rm (stat.)}~\pm~2.3{\rm (syst.)}$ 
events. To estimate the total acceptance of the event selection requirements, we 
use MC samples produced with a leading-order (LO) $Z\gamma$ generator~\cite{baur}, 
passed through a parameterized simulation of the D0~detector. The next-to-leading order (NLO) 
QCD corrections arising from soft gluon radiation and virtual one-loop corrections 
are taken into account through the adjustment of the photon $E_{T}$ spectrum using 
a $K$-factor, estimated using a NLO $Z\gamma$ event generator~\cite{baur_nlo}. 
As we require no jets with $E_T >$~15~GeV to be present in the final state, 
the NLO corrections, integrated over the photon $E_T$ range after the photon $E_T >$~90~GeV requirement, 
are $\approx$~2\% or smaller both for the SM and anomalous $Z\gamma$ production. The NLO 
corrections distribution is fitted with a smooth function, with an uncertainty of $\approx$~5\% 
arising from the fit. The uncertainty on the $K$-factor 
from the jet energy scale and resolution is estimated to be~$\approx$~3\%. Based on this simulation, 
the expected number of events from the SM signal is estimated to be $33.7 \pm 3.4$ events. 
The number of observed events ($N_{\rm obs}$) and the number of predicted events ($N_{\nu\bar\nu\gamma}^{{\rm SM}}$) 
are summarized in Table~\ref{tab:bkg}. 

The $Z\gamma$ cross section multiplied by the branching fraction of $Z \to \nu\bar\nu$ 
is measured to be $32 \pm 9 {\rm (stat.+syst.)} \pm 2 {\rm (lumi.)~fb}$ for the 
photon $E_T~>~90$~GeV, which is in good agreement with the NLO 
cross section of $39 \pm 4$~fb~\cite{baur_nlo}. 
The main contribution to the total uncertainty on the measured cross section is the 
statistical uncertainty on the small number of events in the final sample, and is a factor 
of four to five larger than the individual systematic uncertainties on photon
identification, choice of parton distribution functions (PDF), and kinematic criteria. 
The uncertainty on the theoretical cross section 
comes mainly from the choice of PDF (7\%) and estimation of the NLO $K$-factor (5.5\%). 
To estimate the statistical significance of the measured cross section, we perform 
$10^8$ background-only pseudo-experiments and calculate the p-value as the fraction of 
pseudo-experiments with an estimated cross section above the measured one.
This probability is found to be $3.1\times 10^{-7}$, 
which corresponds to a statistical significance of 5.1 standard deviations (s.d.), 
making this the first observation of the $Z\gamma \to \nu\bar\nu\gamma$ process 
at the Tevatron.

\begin{figure}[htbp]
\begin{center}
\includegraphics[scale=0.35]{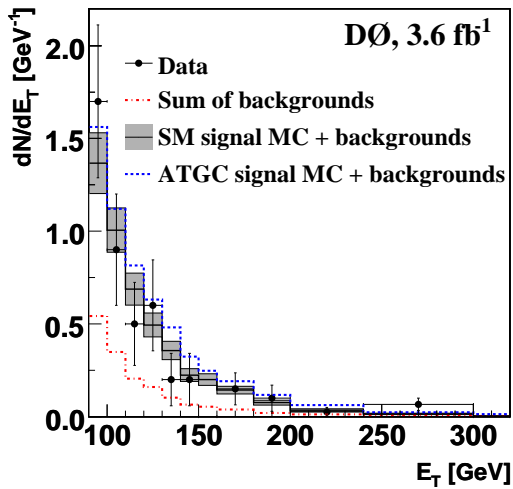}
\caption{\small Photon $E_{T}$ spectrum in data (solid circles), 
sum of backgrounds (dash-dot line), and sum of MC signal and 
background for the SM prediction (solid line) and for 
the ATGC prediction with $h_{30}^{\gamma}=0.09$ and $h_{40}^{\gamma}=0.005$ 
(dashed line). The shaded band corresponds to the $\pm~1$~s.d. total 
uncertainty on the predicted sum of SM signal and background.
\label{fig:ptg_spec}}
\end{center}
\end{figure}

To set limits on the ATGC, we compare the photon $E_T$ spectrum in data with that 
from the sum of expected $Z\gamma$ signal~\cite{baur, baur_nlo} and the background (see Fig.~\ref{fig:ptg_spec}) 
for each pair of couplings for a grid in which $h^{V}_{30}$ runs from -0.12 to 0.12 with 
a step of 0.01, and $h^{V}_{40}$ varies from -0.08 to 0.08 with a step of 0.001.
The MC samples are generated with the LO $Z\gamma$ generator (corrected for the NLO 
effects with an $E_T$-dependent $K$-factor~\cite{baur_nlo}) for the form-factor $\Lambda = 1.5$~TeV.

\begin{figure}[htbp]
\begin{center}
\includegraphics[scale=0.3]{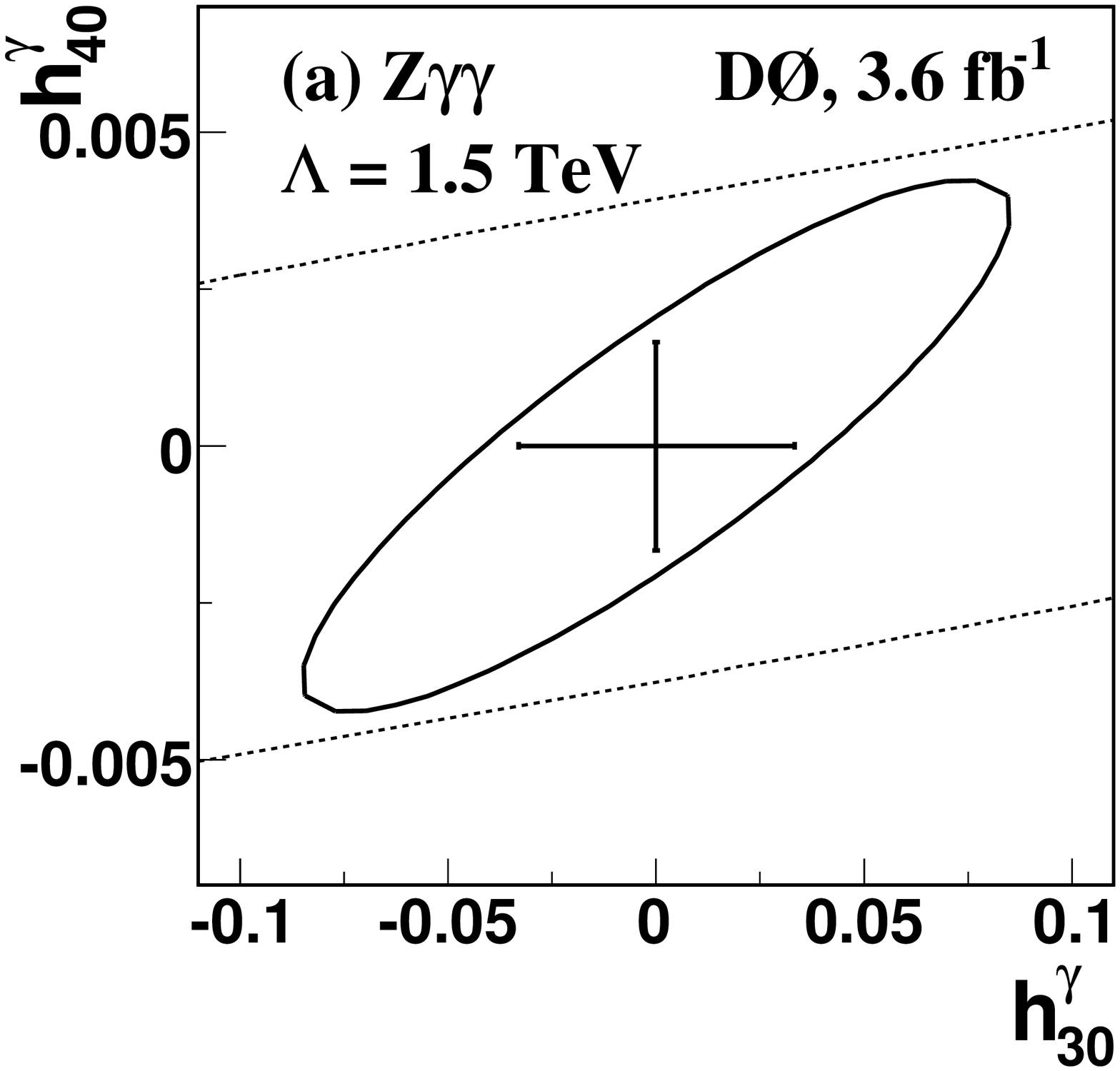}
\includegraphics[scale=0.3]{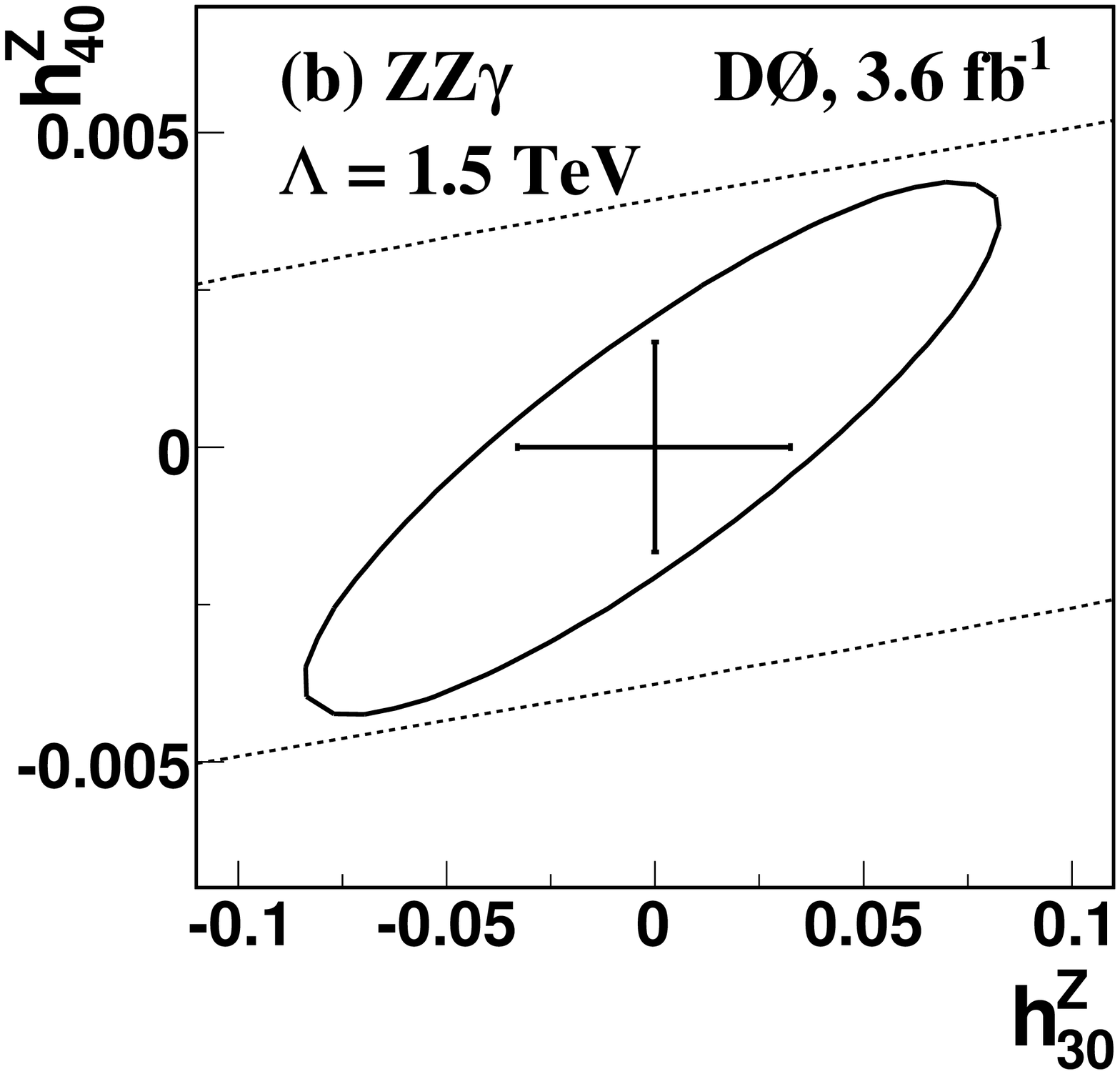}
\caption{\small Two-dimensional bounds (ellipses) at 95\% C.L. on CP-conserving (a) 
$Z\gamma\gamma$ and (b) $ZZ\gamma$ couplings. The crosses represent the one-dimensional bounds
at the 95\% C.L. setting all other couplings to zero. The dashed lines indicate 
the unitarity limits for $\Lambda = 1.5$~TeV.
\label{fig:contours}}
\end{center}
\end{figure}

Assuming Poisson statistics for the signal and Gaussian distribution of all the 
systematic uncertainties on the generated samples and on the backgrounds, we calculate 
the likelihood of the photon $E_{T}$ distribution in data given the prediction for 
hypothesized ATGC. To set limits on any individual ATGC at the 95\% confidence level (C.L.), 
we set the other anomalous couplings to zero.
The resulting limits in the neutrino channel alone are 
$|h_{30}^{\gamma}| < 0.036$, $|h_{40}^{\gamma}| < 0.0019$ 
and $|h_{30}^{Z}| < 0.035$, $|h_{40}^{Z}| < 0.0019$. To further improve the  
sensitivity, we generate the $Z\gamma \to \ell\ell\gamma$ ($\ell = e,~\mu$) MC 
samples for these couplings and $\Lambda = 1.5$~TeV, 
and set limits on ATGC for the 1~fb$^{-1}$ data sample used in the 
previous $Z\gamma$ analysis~\cite{zg_p17}. The combination of all three 
channels yields the most stringent limits on the ATGC set at a hadron 
collider to date: $|h_{30}^{\gamma}| < 0.033$, $|h_{40}^{\gamma}| < 0.0017$ 
and $|h_{30}^{Z}| < 0.033$, $|h_{40}^{Z}| < 0.0017$. This is roughly a factor of three 
improvement over the results published in Ref.~\cite{zg_p17}. The limits on the 
$h_{30}^{Z}$, $h_{40}^{Z}$, and $h_{40}^{\gamma}$ couplings improve on 
the constraints from LEP2, and are the most restrictive to date. The limits on the 
CP-violating couplings $h_{10}^V$ and $h_{20}^V$ are, within the precision of 
this measurement, the same as the limits on $h_{30}^V$ and $h_{40}^V$, respectively. Hence, we 
can constrain the strength of the couplings but not the phase. 
As the described method is sensitive only to the magnitude and the relative sign between couplings,
the one- and two-dimensional limits are symmetric with respect to the SM coupling under 
simultaneous exchange of all signs. The 95\% C.L. 
one-dimensional limits and two-dimensional contours are shown in Figs.~\ref{fig:contours}a and~\ref{fig:contours}b 
for the CP-conserving $Z\gamma\gamma$ and $ZZ\gamma$ couplings, respectively.

In summary, we observe 51 $\nu\bar\nu\gamma$ candidates with $17.3~\pm~0.6{\rm (stat.)}~\pm~2.3{\rm (syst.)}$ 
background events using 3.6~fb$^{-1}$ of data collected with the D0~detector at the Tevatron.
We measure the most precise $Z\gamma \to \nu\bar\nu\gamma$ cross section to date at a hadron collider of 
$32 \pm 9{\rm (stat. + syst.)} \pm 2{\rm (lumi.)}$~fb for the photon $E_{T} > 90$~GeV, in agreement with the SM 
prediction of $39 \pm 4$~fb~\cite{baur_nlo}. The statistical significance of this measurement is 
5.1~s.d., making it the first observation of the $Z\gamma \to \nu\bar\nu\gamma$ process 
at the Tevatron. We set the most restrictive limits on the real parts of the anomalous trilinear 
gauge couplings at hadron colliders at the 95$\%$ C.L. of 
$|h_{30}^{\gamma}| < 0.033$, $|h_{40}^{\gamma}| < 0.0017$ 
and $|h_{30}^{Z}| < 0.033$, $|h_{40}^{Z}| < 0.0017$. Three of these limits are world's best to date. 
These limits approach the range of expectations for the contributions due to one-loop diagrams in the SM
~\cite{Gounaris_2002za,zzg-theory}.

%
We thank the staffs at Fermilab and collaborating institutions, 
and acknowledge support from the 
DOE and NSF (USA);
CEA and CNRS/IN2P3 (France);
FASI, Rosatom and RFBR (Russia);
CNPq, FAPERJ, FAPESP and FUNDUNESP (Brazil);
DAE and DST (India);
Colciencias (Colombia);
CONACyT (Mexico);
KRF and KOSEF (Korea);
CONICET and UBACyT (Argentina);
FOM (The Netherlands);
STFC (United Kingdom);
MSMT and GACR (Czech Republic);
CRC Program, CFI, NSERC and WestGrid Project (Canada);
BMBF and DFG (Germany);
SFI (Ireland);
The Swedish Research Council (Sweden);
CAS and CNSF (China);
and the
Alexander von Humboldt Foundation (Germany).
%


\begin{thebibliography}{99}
\bibitem{eta} Pseudorapidity, $\eta$, is defined as $\eta = -\text{ln}\left[\text{tan}\left(\frac{\theta}{2}\right)\right]$, where $\theta$ is the polar angle measured with respect to the proton beam direction.
\bibitem{dca} Distance of closest approach is defined as the shortest distance between the 
particle's trajectory and the $z$-axis in the $x-y$ plane.
\end{thebibliography}
\end{document}